\input{aipcheck}

\documentclass[
    ,final            % use final for the camera ready runs
,numberedheadings % uncomment this option for numbered sections
  ]
  {aipproc}

\layoutstyle{8x11double}

\begin{document}

\title{Searching for Neutrino Radio Flashes from the Moon with LOFAR}

%Neutrino, muon, pion, and other elementary particles; cosmic rays 

\author{Stijn Buitink}{
  address={Kernfysisch Versneller Instituut, 9747 AA Groningen, The Netherlands},
  altaddress={Dept. of Astrophysics/IMAPP, Radboud University Nijmegen, 6500 GL Nijmegen, The Netherlands}
}

\author{Arthur Corstanje}{
  address={Dept. of Astrophysics/IMAPP, Radboud University Nijmegen, 6500 GL Nijmegen, The Netherlands}
}

\author{Emilio Enriquez}{
  address={Dept. of Astrophysics/IMAPP, Radboud University Nijmegen, 6500 GL Nijmegen, The Netherlands}
}

\author{Heino Falcke}{
  address={Dept. of Astrophysics/IMAPP, Radboud University Nijmegen, 6500 GL Nijmegen, The Netherlands},
  altaddress={Netherlands Institute for Radio Astronomy (ASTRON), 7990 AA Dwingeloo, The Netherlands}
}

\author{Wilfred Frieswijk}{
  address={Netherlands Institute for Radio Astronomy (ASTRON), 7990 AA Dwingeloo, The Netherlands}
}

\author{J\"org H\"orandel}{
  address={Dept. of Astrophysics/IMAPP, Radboud University Nijmegen, 6500 GL Nijmegen, The Netherlands},
  altaddress={Nikhef, Science Park Amsterdam, 1098 XG Amsterdam, The Netherlands}
}

\author{Maaijke Mevius}{
  address={Netherlands Institute for Radio Astronomy (ASTRON), 7990 AA Dwingeloo, The Netherlands}
}

\author{Anna Nelles}{
  address={Dept. of Astrophysics/IMAPP, Radboud University Nijmegen, 6500 GL Nijmegen, The Netherlands},
  altaddress={Nikhef, Science Park Amsterdam, 1098 XG Amsterdam, The Netherlands}
}

\author{Satyendra Thoudam}{
  address={Dept. of Astrophysics/IMAPP, Radboud University Nijmegen, 6500 GL Nijmegen, The Netherlands}
}

\author{Pim Schellart}{
  address={Dept. of Astrophysics/IMAPP, Radboud University Nijmegen, 6500 GL Nijmegen, The Netherlands}
}

\author{Olaf Scholten}{
  address={Kernfysisch Versneller Instituut, 9747 AA Groningen, The Netherlands}
}

\author{Sander ter Veen}{
  address={Dept. of Astrophysics/IMAPP, Radboud University Nijmegen, 6500 GL Nijmegen, The Netherlands}
}

\author{Martin van den Akker}{
  address={Dept. of Astrophysics/IMAPP, Radboud University Nijmegen, 6500 GL Nijmegen, The Netherlands}
}

\author{the LOFAR collaboration}{
  address={Netherlands Institute for Radio Astronomy (ASTRON), 7990 AA Dwingeloo, The Netherlands}
}

\begin{abstract}
Ultra-high-energy neutrinos and cosmic rays produce short radio flashes through the Askaryan effect when they impact on the Moon.
Earthbound radio telescopes can search the Lunar surface for these signals.
A new generation of low-frequency, digital radio arrays, spearheaded by LOFAR, will allow for searches with unprecedented sensitivity. In the first stage of the NuMoon project, low-frequency observations were carried out with the Westerbork Synthesis Radio Telescope, leading to the most stringent limit on the cosmic neutrino flux above $10^{23}$ eV. With LOFAR we will be able to reach a sensitivity of over an order of magnitude better and to decrease the threshold energy.
\end{abstract}

\maketitle

%%%%%%%%%%%%%%%%%%%%%%%%%%%%%%%%%%%%%%%%%%%%
%% MAINMATTER
%%%%%%%%%%%%%%%%%%%%%%%%%%%%%%%%%%%%%%%%%%%%

\section{Introduction}
A century after their discovery, cosmic rays (CRs) still represent one of the unsolved questions in astrophysics. While modern experiments like the Pierre Auger Observatory (Auger) \cite{PAO_correlation}, HiRes \cite{HiRes_GZK} and the Telescope Array (TA) \cite{TA} can efficiently detect ultra-high-energy (UHE) CRs, two complications remain that make it hard to identify their sources. First, the trajectories of CRs are bent in the (inter-)Galactic magnetic fields, so the arrival directions of the CRs at Earth do not align with the positions of their sources. This deflection depends on the charge of the CRs, and therefore on their composition. 
An Auger study of possible correlation with AGNs \cite{PAO_correlation} disfavors CR isotropy above $6\cdot 10^{19}$ eV.
However, composition studies by Auger show a trend towards a heavier composition at the highest energies \cite{PAO_composition}, which suggests large separations between CR arrival directions and sources. In contrast, HiRes data does not reject isotropy and suggests a light CR composition \cite{HiRes_composition}. 
A second complication is that CRs above an energy of $4\cdot 10^{19}$~eV will interact with the cosmic microwave background (CMB). In these Greisen-Zatsepin-Kuzmin (GZK) interactions pions are created, which produce neutrinos when they decay \cite{g66,Zatsepin}. A steepening of the CR spectrum at this energy has been found experimentally \cite{PAO_GZK, HiRes_GZK}. The attenuation length of protons due to the GZK effect is $\sim 50$ Mpc. Sources outside this radius cannot be found by studying UHE CRs directly. 
The detection of UHE neutrinos that are produced in GZK interactions or directly in the source is an attractive alternative to find these sources. Neutrinos are not affected by magnetic fields, and they can travel over cosmic distances almost unattenuated.

It was first suggested by Dagkesamanskii and Zheleznykh \cite{dz89} to use the Moon as a detector for UHE neutrinos and CRs. At energies above several PeV the Moon is opaque to neutrinos. Neutrinos impinging on the Moon will interact below the lunar surface. In a charged current interaction $\sim20$\% of the energy is converted into a hadronic cascade. The leptons that are created are in principle detectable, though in the case of muons and taus the relevant cross-sections at UHE are highly uncertain. Electrons initiate an electromagnetic shower which is hard to detect with the lunar Cherenkov technique because it is elongated due to the LPM effect \cite{LPM}. However, at sufficiently high energy, the length and depth of the cascade is reduced because of the increase of the photonuclear and electronuclear cross sections \cite{Gerhardt}.
Particle cascades beneath the lunar surface can be detected by the radiation they emit through the Askaryan effect \cite{a62}.

GLUE  \cite{glue} was one of the pioneering experiments at high frequencies ($>$1 GHz). Presently, LUNASKA \cite{lunaska}  performs high frequency measurements with ATCA and the Parkes radio telescope. At the Expanded Very Large Array,  RESUN \cite{RESUN} performs 1.4 GHz lunar observations.

The NuMoon project observes in a low frequency window which offers an optimal detection probability at the highest energies. In the first phase, observations were performed with the Westerbork Synthesis Radio Telescope (WSRT) leading to the most constraining limit on the UHE neutrino flux above $10^{23}$ eV \cite{NuMoon-PRL, NuMoonAA}. Presently, we are preparing observations with the Low Frequency Array (LOFAR) \cite{lofar}.

\section{Radio emission from lunar particle cascades}
Neutrinos and CRs impacting on the Moon initiate cascades below the surface, which develop a time-varying negative charge excess propagating at the speed of light in vacuum, and thereby produce radio emission \cite{a62}. Although the radiation is most intense at the Cherenkov angle, it can be misleading to think of it in terms of pure Cherenkov radiation. The rise and decay of a moving charge itself would also produce radiation if the index of refraction of the medium were unity \citep{terVeen, J10}. Actually, at low frequencies, when the wavelength is larger than the cascade length, the radiation is coherent at angles far from the Cherenkov angle \citep{ZHS92}. 

The spread of the radiation around the Cherenkov angle is of large importance for the detectability of the radio pulse. To observe the pulse on Earth, it needs to be able to escape the Moon. Since the Cherenkov angle is equal to the angle of total internal reflection on the surface, the high frequency component can only leave the Moon if the shower points upwards to the surface. This is only the case when a particle enters the Moon from behind, near the lunar rim. The low frequency component, on the other hand, is spread out over a much larger opening angle, and part of the radiation can escape the Moon even if the cascade is directed downwards into the Moon. At $\sim$150 MHz the whole visible surface of the Moon contributes to the effective detector volume \cite{Scholten06}.

We simulated the distribution of observable pulses over the lunar surface for an isotropic neutrino flux with an energy spectrum $E^{-1}$ in the range $10^{21}-10^{24}$~eV. 
We assume that 20\% of the neutrino energy is deposited in a hadronic cascade for all flavors.
The neutrino cross sections \cite{cc} are extrapolated to high energies for a 500~m deep layer of regolith, and the radio pulse parametrization of Alvarez-Mu\~niz et al. \cite{amparam} is used for the hadronic showers. 
As a mean value for the attenuation length for the radiated power we have taken $\lambda = (9/\nu[\mathrm{GHz}])$~m. Surface roughness is not included since it does not play an important role at low frequencies \cite{Scholten06}.

\begin{figure}
\caption{\label{geoplot} Geometry:  $\theta_{\nu}$ is defined as the angle between the direction of the neutrino and the observer; $b$ is the relative distance of the point where the pulse emerges from the lunar surface to the center of the face of the Moon.}
\includegraphics[width=0.99\columnwidth]{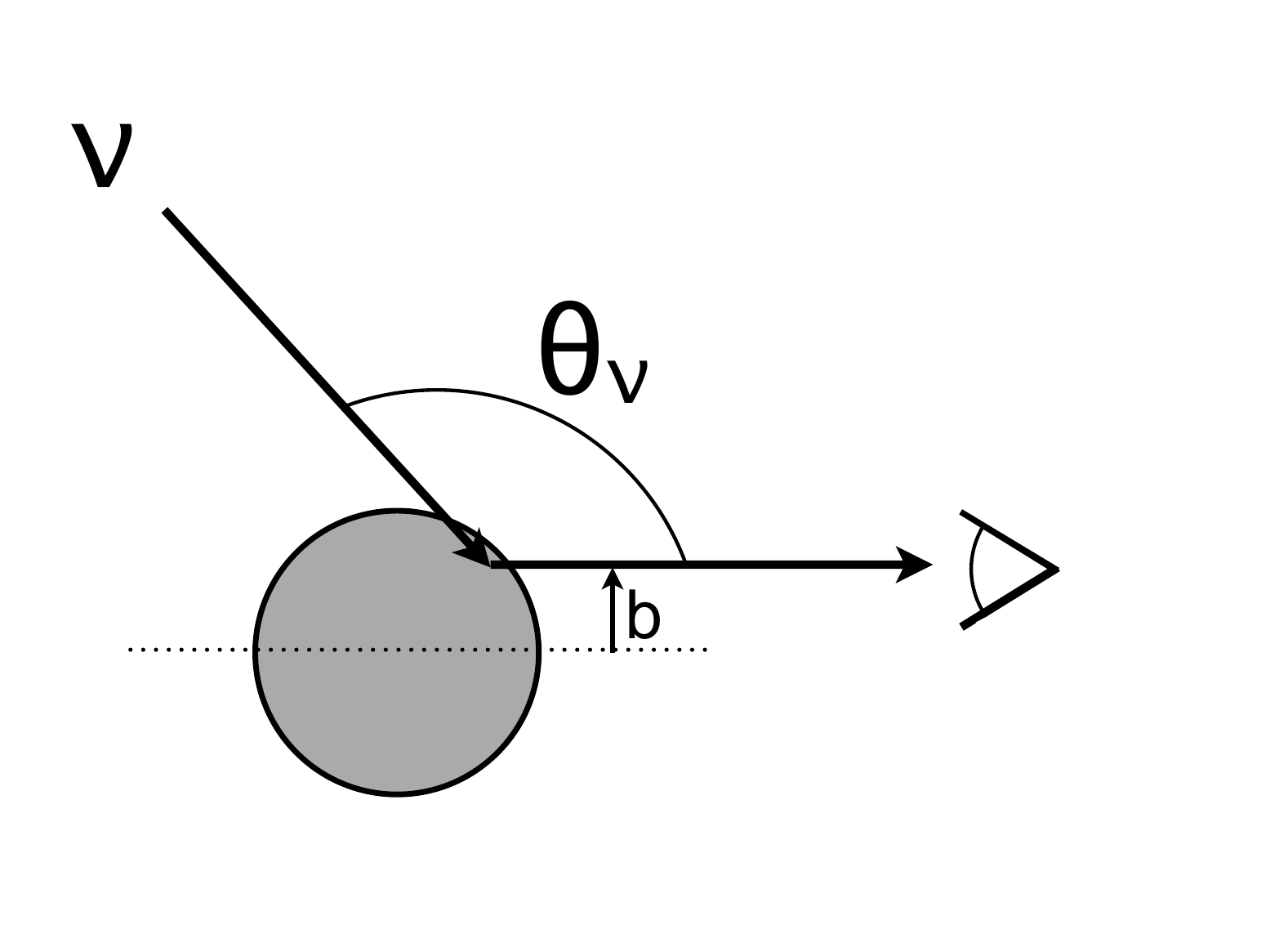}
\end{figure}

\begin{figure}
\caption{\label{simplot} Distribution of radio pulses with power greater than 500~Jy as a function of energy and $b$.}
\includegraphics[width=0.99\columnwidth]{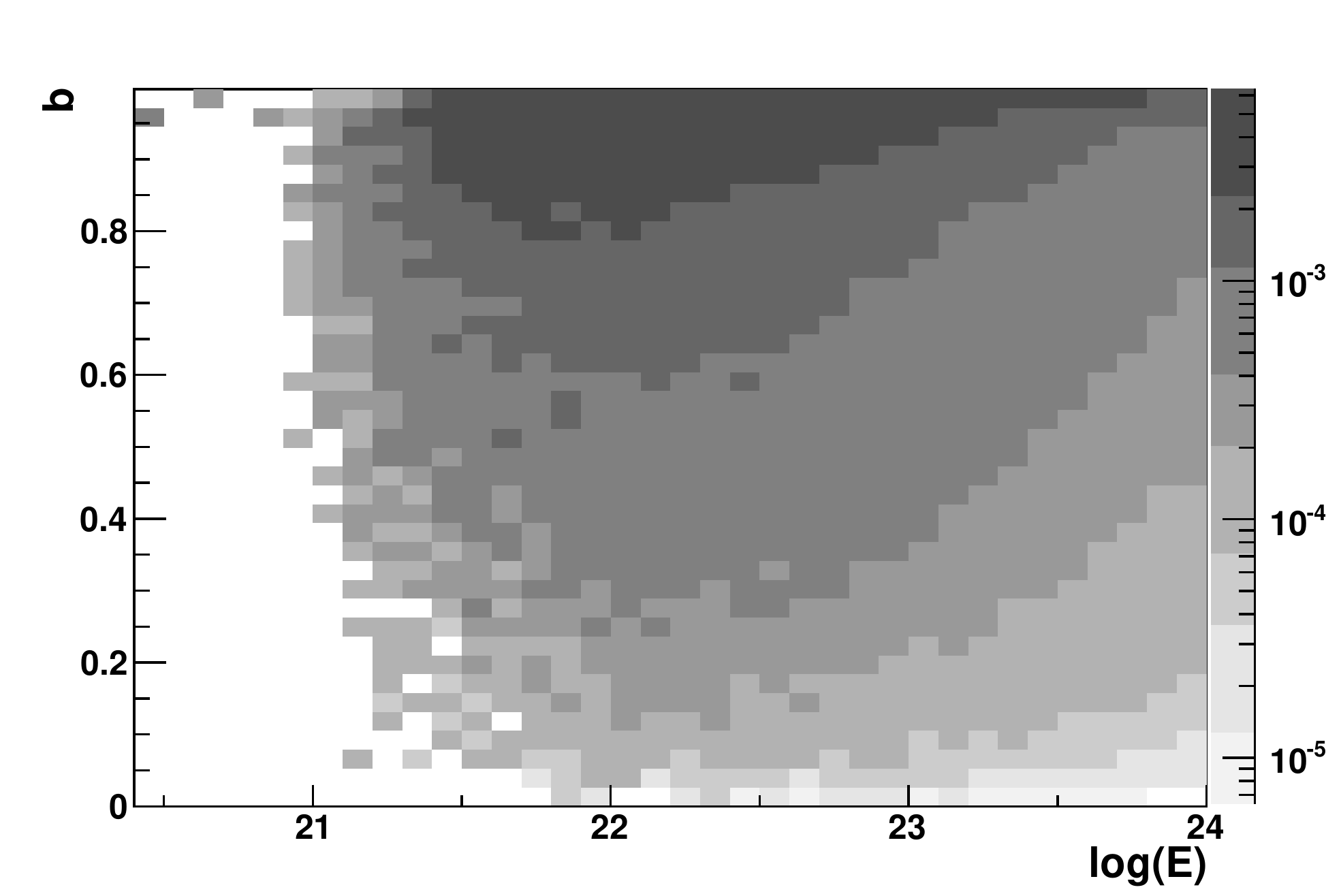}
\end{figure}

\begin{figure}
\caption{\label{simplot2} Distribution of radio pulses with power greater than 500~Jy as a function of $\theta_{\nu}$ and $b$.}
\includegraphics[width=0.99\columnwidth]{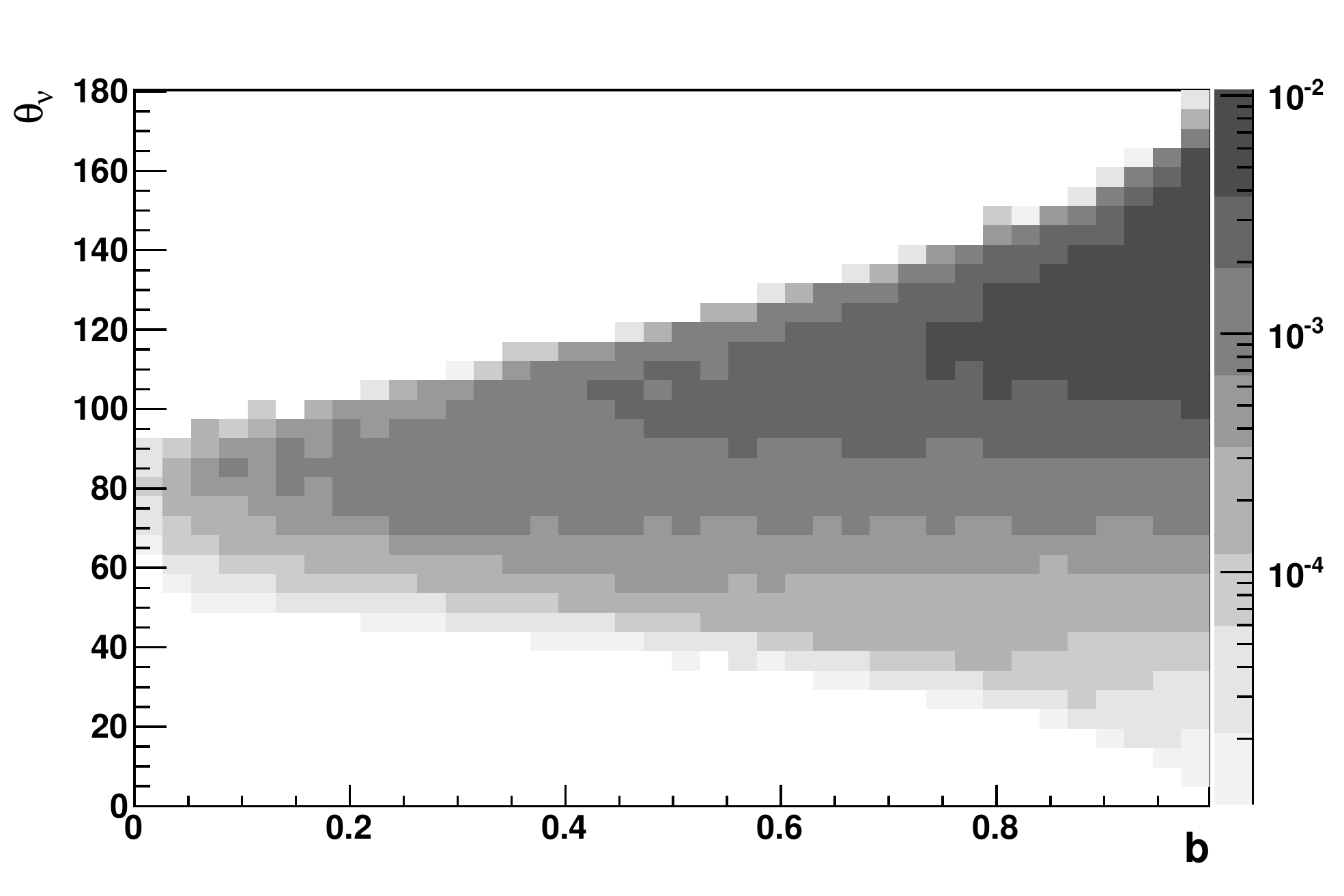}
\end{figure}

The geometry definitions used here are sketched in Fig.~\ref{geoplot}: $\theta_{\nu}$ is defined as the angle between the direction of the neutrino and the observer; $b$ is the relative distanceof the point where the pulse emerges from the lunar surface to the center of the face of the Moon ($b=1$ corresponds to the rim of the Moon).

In Fig.~\ref{simplot} the distribution of radio pulses above a threshold of 500~Jy in the 110$-$190 MHz  band is plotted as a function of neutrino energy and $b$. The energy spectrum has been reweighted to $E^{-2}$ and the total observed flux has been normalized to unity. The color scale thus represents the relative probabilities of observing an event in different parts of the parameter space. The threshold energy for detection is $\sim 10^{21}$~eV and at one decade in energy higher the whole surface of the Moon contributes to the detector volume. Even at higher energies events are more likely to occur close to the rim. This is mainly a projection effect: the high-$b$ region represents a very large part of the visible half of the lunar sphere.

In Fig.~\ref{simplot2} the same set of events is now plotted as a function of $\theta_{\nu}$ and $b$. There is a strong clustering of events towards the rim of the Moon with an arrival angle $\theta_{\nu}$ in the range $110^\circ - 150^\circ$. Neutrinos from $50^\circ - 90^\circ$ can produce observable pulses on the complete lunar surface, while for even smaller values of  $\theta_{\nu}$, observable events are rare. Hence, localization of a low-frequency radio pulse on the Moon provides constraints on the arrival direction of the neutrino.

\section{NuMoon with LOFAR}
LOFAR \cite{lofar} is a new kind of radio telescope consisting of thousands of simple omni-directional antennas that takes advantage of recent developments in fast electronics. The received radio signal is digitized at each antenna and sent to a central processor where directed beams are synthesized by applying the appropriate time delays. Because the antennas themselves contain no moving components they are relatively cheap and can be constructed in large amounts. Another great advantage of synthesizing beams through software is that multiple beams in different directions can be in operation at the same time.

\begin{figure}
\caption{\label{fig:superterp}Arial view of the LOFAR core. The island, called the \emph{superterp} contains 6 of the 24 core stations. It is divided in larger and smaller patches which house the LBA and HBA antennas respectively.}
\includegraphics[width=0.8\columnwidth]{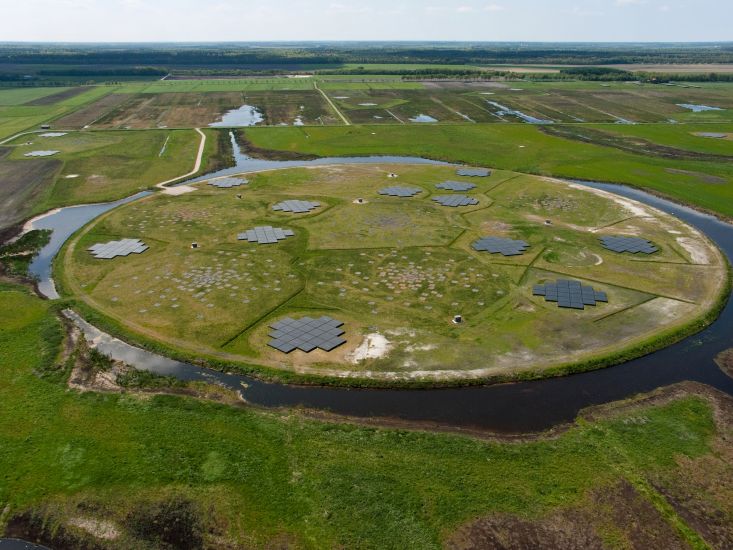}
\end{figure}

Each LOFAR station consists of 96 Low Band Antennas (LBA), which operate in the 10$-$80~MHz range, and 48 High Band Antennas (HBA), which cover the 110$-$240~MHz band. Currently, the array consists of 24 core stations, concentrated in an area of $\sim$12 km$^2$, 9 remote stations, distributed over the northern part of the Netherlands with baselines up to 80~km from the core, and 8 international stations.  

For the NuMoon experiment we will use the HBA antennas operating in the $110-190$~MHz window. The beamforming is done in two stages because of limited bandwidth between stations and the central processor. At each station a beam of a few degrees opening angle is formed in the direction of the Moon. This signal is transported to the central processor where it is first corrected for ionospheric dispersion. At LOFAR frequencies, a typical ionospheric electron content of 10 TEC-units (TECU=10$^{16}$ electron/m$^{2}$) causes a dispersion which spreads a bandwidth-limited signal pulse over hundreds of nanoseconds \cite{mevius}. In order to be able to perform a pulse search the signal needs to be de-dispersed with an accuracy of 1 TECU. An interesting approach is measuring the absolute TEC value via the Faraday rotation of polarized light due to ionospheric plasma and the Earth magnetic field. We are currently investigating the possibility to use the polarized light of the rim of the Moon for this purpose.

In the next step, the de-dispersed station beams are combined to form array beams of $\sim 0.1$ degree width. Up to 50 such beams can be synthesized simultaneously to cover the complete surface of the Moon. Each of these beams will be searched in real-time for short pulses. When a candidate pulse is found in one of the beams a coincidence check will be performed. A radio pulse originating from a certain location on the Moon is strong in only one of the array beams (although neighboring beams may also receive a signal, depending on beam shape and sidelobes). This is a powerful vetoing technique that removes pulses from strong noise sources near one of the stations: a very strong noise pulse in one of the station beams will cause a spike in all the array beams.

Each of the LOFAR antennas is connected to a ring buffer that stores the last five seconds of raw data. When a pulse is found at the central processor that passes the anti-coincidence veto, a trigger will be sent back to the stations and the buffers are read out and permanently stored. This allows us to use the full resolution, full bandwidth raw time-series data of all antennas for offline analysis \cite{kalpana}.   

\section{Sensitivity}

\begin{figure}
\caption{\label{fig:nulimit}All-flavour neutrino upper limits and future sensitivities of several experiments. See text for details.}
\includegraphics[width=0.99\textwidth]{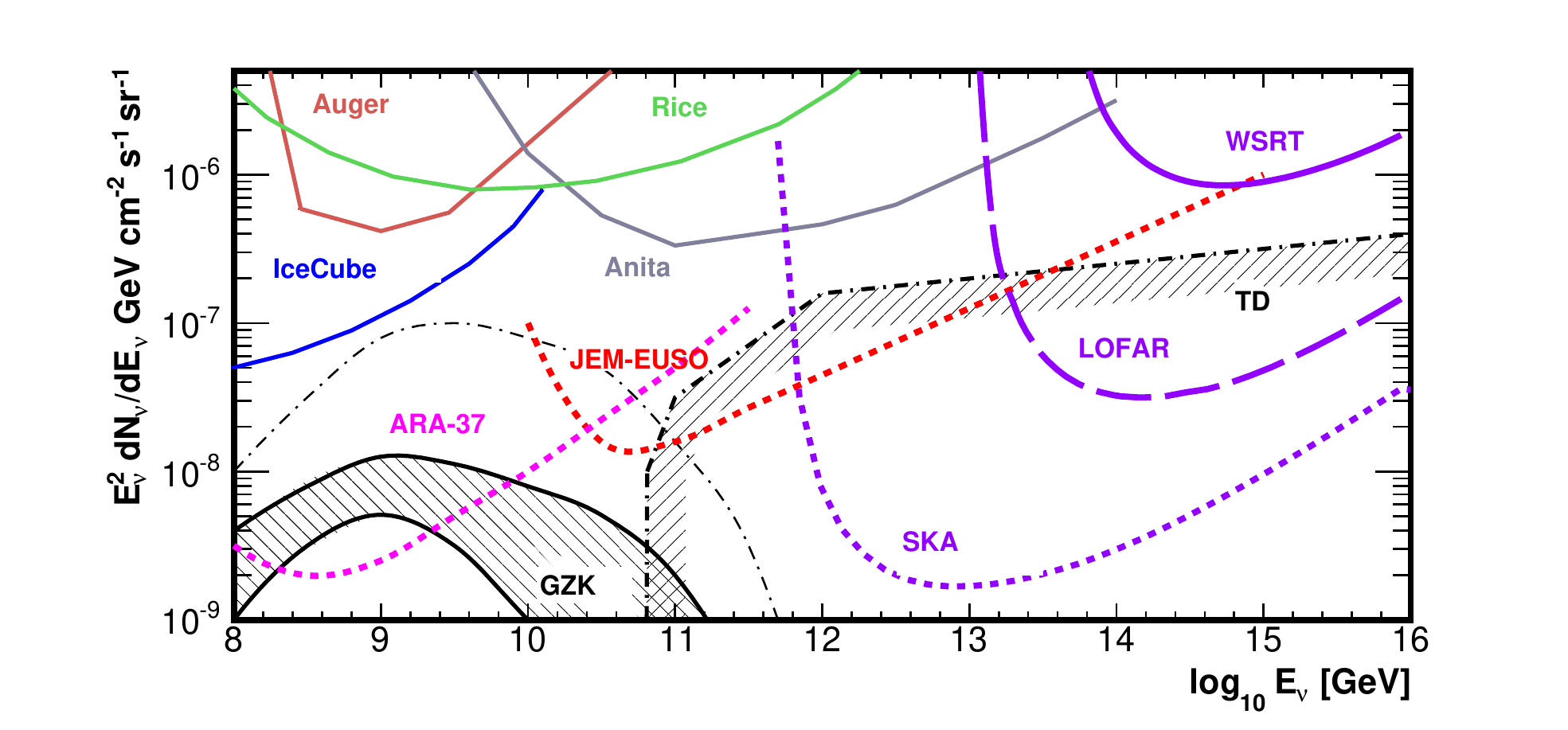}
\end{figure}

\begin{figure}
\caption{\label{fig:crlimit}CR flux limit of WSRT, as well as expected sensitivities of LOFAR and SKA. The dotted line indicates a constant particle flux (cm$^{-2}$sr$^{-1}$s$^{-1}$) corresponding to the particle flux of the highest-energy data point. A flux above this line would have been found by Auger.}
\includegraphics[width=0.99\columnwidth]{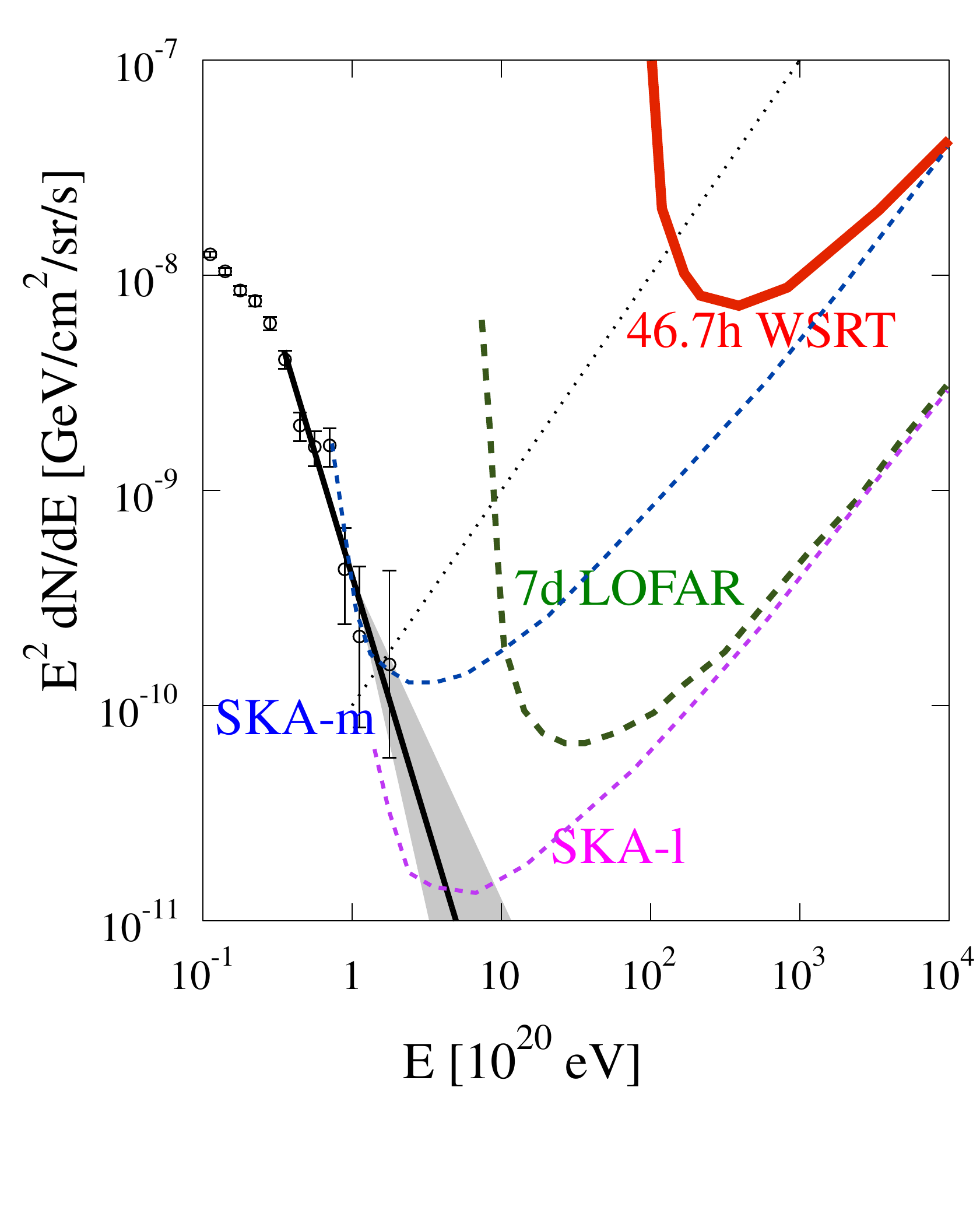}
\end{figure}

In the past years, much progress had been made in improving the sensitivity to cosmic neutrinos over a large range of energy, with a large range of techniques. In Fig.~\ref{fig:nulimit} existing upper limits on the all-flavor neutrino flux are plotted in solid lines. The most stringent limits are set by IceCube \cite{ICnu}, ANITA \cite{ANITA} and the WSRT \cite{NuMoon-PRL} at increasingly larger energies. Also plotted are limits published by Rice \cite{rice} and Auger \cite{PAO_nu}.  

The expected sensitivity that is reached with one week of LOFAR data will improve the previous WSRT limit by an order of magnitude, while also decreasing the threshold detection energy. In the future, the Square Kilometer Array (SKA) \cite{SKA} will provide an even better sensitivity using the same techniques as LOFAR. At lower energies, better sensitivities will be reached by future experiments ARA \cite{ARA} and ARIANNA \cite{ARIANNA} (not plotted), both searching for radio flashes of neutrinos in ice, and the JEM-EUSO satellite \cite{JEM}. 

The flux of GZK neutrinos is sensitive to various unknown parameters including source evolution, CR composition and source energy cutoff. Figure \ref{fig:nulimit} contains predictions computed by \cite{KAO10} based on varying input parameters. The shaded band represents a region which contains a large part of the possible models. The model that produces the largest neutrino flux is indicated by dash-dotted line.

UHE neutrinos can also be created in the decay of supermassive particles (top-down scenarios). The TD-curve in Fig.~\ref{fig:nulimit} delimits the parameter space in which neutrinos may be found originating from moduli --- weakly coupled scalar particles predicted in supersymmetric theories --- from the kinks of cosmic string loops \citep{BerTD,LSTD}.

A cascade induced by a CR will be initiated just below the lunar surface in contrast to neutrino induced showers. Therefore, the radio signal suffers less attenuation. However,  the question rises whether or not the radiative field can properly develop in a volume that is smaller than the wavelength of the radiation. A detailed derivation \citep{terVeen, J10} shows that such a formation zone is not needed. This becomes evident when one realizes that a system of growing and decaying current would also radiate if it were to develop in a vacuum. Hence, the dielectric medium is not essential for the formation of the radiative field, although it certainly affects it. In Fig.~\ref{fig:crlimit} the WSRT upper limit on CR flux \citep{terVeen} is plotted, as well as expected sensitivities for LOFAR and SKA (low band and middle band). The data points represent the high end of the CR spectrum as measured by Auger \cite{PAO_GZK}.

\section{Conclusions}
The Numoon experiment uses the Moon as a ultra-high-energy cosmic-ray and neutrino detector. It is different from other lunar Askaryan experiments as it searches for radio flashes at lower frequencies ($110-190$~MHz). This has the advantage that a much higher sensitivity can be reached, although the threshold energy is slightly lower at high frequencies (compare for example the sensitivities of the SKA low and middle band in Fig.~\ref{fig:crlimit}) . Observations with the WSRT have led to the most stringent upper limits above $10^{23}$ eV. Preparations are now in progress to perform measurements with LOFAR. The expected sensitivity is over an order of magnitude better than previous measurements. In the future, the SKA will make another big leap in sensitivity using the same techniques. 

%%%%%%%%%%%%%%%%%%%%%%%%%%%%%%%%%%%%%%%%%%%%%%%%
%% BACKMATTER
%%%%%%%%%%%%%%%%%%%%%%%%%%%%%%%%%%%%%%%%%%%%%%%%

\begin{theacknowledgments}
This work was supported by the Netherlands Organization for Scientific Research (NWO), VENI grant 639-041-130, the Stichting
voor Fundamenteel Onderzoek der Materie (FOM),
the Samenwerkingsverband Noord-Nederland (SNN)
and the Netherlands Research School for Astronomy (NOVA).
LOFAR, the Low Frequency Array designed and constructed by ASTRON, has facilities in several countries, that are owned by various parties (each with their own funding sources), and that are collectively operated by the International LOFAR Telescope (ILT) foundation under a joint scientific policy. 
\end{theacknowledgments}

%%%%%%%%%%%%%%%%%%%%%%%%%%%%%%%%%%%%%%%%%%%%%%%%
%% The bibliography can be prepared using the BibTeX program or
%% manually.
%%
%% The code below assumes that BibTeX is used.  If the bibliography is
%% produced without BibTeX comment out the following lines and see the
%% aipguide.pdf for further information.
%%
%% For your convenience a manually coded example is appended
%% after the \end{document}
%%%%%%%%%%%%%%%%%%%%%%%%%%%%%%%%%%%%%%%%%%%%%%%%

%%%%%%%%%%%%%%%%%%%%%%%%%%%%%%%%%%%%%%%%%%%%%%%%
%% You may have to change the BibTeX style below, depending on your
%% setup or preferences.
%%
%%
%% For The AIP proceedings layouts use either
%%%%%%%%%%%%%%%%%%%%%%%%%%%%%%%%%%%%%%%%%%%%

\bibliographystyle{aipproc}   % if natbib is available
%\bibliographystyle{aipprocl} % if natbib is missing

%%%%%%%%%%%%%%%%%%%%%%%%%%%%%%%%%%%%%%%%%%%
%% You probably want to use your own bibtex database here
%%%%%%%%%%%%%%%%%%%%%%%%%%%%%%%%%%%%%%%%%%%
%\bibliography{sample}

%%%%%%%%%%%%%%%%%%%%%%%%%%%%%%%%%%%%%%%%%%%
%% Just a reminder that you may have to run bibtex
%% All of it up to \end{document} can be removed
%% if you don't like the warning.
%%%%%%%%%%%%%%%%%%%%%%%%%%%%%%%%%%%%%%%%%%%
%\IfFileExists{\jobname.bbl}{}
% {\typeout{}
 % \typeout{******************************************}
 % \typeout{** Please run "bibtex \jobname" to optain}
 % \typeout{** the bibliography and then re-run LaTeX}
 % \typeout{** twice to fix the references!}
 % \typeout{******************************************}
 % \typeout{}
% }

%%%%%%%%%%%%%%%%%%%%%%%%%%%%%%%%%%%%%%%%%%%
%% The following lines show an example how to produce a bibliography
%% without the help of the BibTeX program. This could be used instead
%% of the above.
%%%%%%%%%%%%%%%%%%%%%%%%%%%%%%%%%%%%%%%%%%%

\end{document}